\documentclass[]{jfm}

\usepackage{graphicx}
\usepackage{newtxtext}
\usepackage{newtxmath}
\usepackage{natbib}
\usepackage{hyperref}
\hypersetup{
    colorlinks = true,
    urlcolor   = blue,
    citecolor  = black,
}

\newcommand{\bu}{\boldsymbol{u}}
\newcommand{\buL}{\bu^\mathrm{L}}
\newcommand{\buS}{\bu^\mathrm{S}}

\newcommand{\uS}{u^{\mathrm{S}}}
\newcommand{\pE}{p^\mathrm{E}}
\renewcommand{\d}{\partial}
\newcommand{\ee}{\mathrm{e}}

\newcommand{\defn}{\equiv}
\newcommand{\bxh}{\boldsymbol{\hat x}}
\newcommand{\byh}{\boldsymbol{\hat y}}
\newcommand{\bzh}{\boldsymbol{\hat z}}
\newcommand{\lap}{\triangle}
\newcommand{\beq}{\begin{equation}}
\newcommand{\eeq}{\end{equation}}
\newcommand{\com}{\, ,}
\newcommand{\per}{\, .}

\newcommand{\di}{\, \mathrm{d}}
\newcommand{\J}{\mathrm{J}}

\title{Transition to turbulence in wind-drift layers}

\author{
    Gregory LeClaire Wagner, \aff{a} \corresp{Gregory Wagner, wagner.greg@gmail.com} 
    Nick Pizzo,              \aff{b,c} 
    Luc Lenain,              \aff{c} 
    Fabrice Veron            \aff{d} 
}

\affiliation{
    \aff{a}{Massachusetts Institute of Technology}\\
    \aff{b}{University of Rhode Island Graduate School of Oceanography}\\
    \aff{c}{Scripps Institution of Oceanography, University of California, San Diego}
    \aff{d}{University of Delaware}\\
}

\begin{document}
\maketitle

\begin{abstract}
A light breeze rising over calm water initiates an intricate chain of events that culminates in a centimeters-deep turbulent shear layer capped by gravity-capillary ripples.
At first, viscous stress accelerates a laminar wind-drift layer until small surface ripples appear.
Then a second ``wave-catalyzed'' instability grows in the wind-drift layer, before sharpening into along-wind jets and downwelling plumes, and finally devolving into three-dimensional turbulence.
This paper elucidates the evolution of wind-drift layers after ripple inception using wave-averaged numerical simulations with a random initial condition and a constant-amplitude representation of the incipient surface ripples.
Our model reproduces qualitative aspects of laboratory measurements similar those reported by \cite{veron2001experiments}, validating the wave-averaged approach.
But we also find that our results are disturbingly sensitive to the amplitude of the prescribed surface wave field, raising the question whether wave-averaged models are truly ``predictive'' if they do not also describe the evolution of the coupled evolution of the surface waves together with the flow beneath.
\end{abstract}

\section{Introduction}

The appearance of surface ripples beneath gusts of wind is an everyday experience on the water, belying a surprisingly intricate chain of events unfolding beneath the surface.
There, an accelerating wind-drift layer breeds two instabilities in sequence: first the surface instability that generates ripples, followed by a subsurface instability whose growth, finite amplitude saturation, and destabilization to three-dimensional perturbations ultimately gives way to persistent turbulence in the centimeters-thick wind-drift layer.

This ubiquitous transition-to-turbulence scenario was observed in a series of laboratory experiments reported by \cite{melville1998laboratory}, \cite{veron1999pulse} and \cite{veron2001experiments} that subjected an initially quiescent wave tank to a turbulent airflow rapidly accelerated from rest to a constant airspeed.
One of \cite{veron2001experiments}'s key results is that both the generation of surface ripples and the transition to turbulence are suppressed by surfactant layered on the water surface.
The surfactant experiment proves that ripples are intrinsic to the second slow instability implicated in the turbulent transition of the wind-drift layer under typical conditions (disproving \cite{handler2001thermal}'s hypothesis, repeated by \cite{thorpe2004langmuir}, that the transition to turbulence is convective.)

Motivated by \cite{veron2001experiments}'s experimental results, we propose a wave-averaged model based on the ``Craik-Leibovich'' (CL) Navier-Stokes equations \citep{craik1976rational} for the development of wind-drift layers --- eventually leading to a transition to turbulence --- after the appearance of capillary ripples.
We focus narrowly on a comparison with a new laboratory experiment similar to those reported by \cite{veron2001experiments} and described in section~\ref{lab}.
The predictions of our wave-averaged model are developed in section~\ref{theory}.
We mix analytical results from \cite{veron2001experiments} on the initial laminar developments with a linear instability analysis of the wind-drift layer just after ripple inception, and with numerical simulations of nonlinear development of the second slow instability from ripple inception to fully developed wind-drift turbulence.

We have two goals: first, we seek a more detailed understanding of the wind-drifted transition to turbulence.
Second, we seek to validate the wave-averaged Craik-Leibovich (CL) momentum equation, which is central to parameterization of ocean surface boundary layer turbulence (see for example \cite{d2014quantifying, harcourt2015improved, reichl2019parameterization}).
Toward this second goal we make some progress and find that our CL-based model qualitatively replicates the laboratory measurements --- most strikingly during the transition to turbulence depicted in figure~\ref{fig:self-sharpening}.
Yet we also find our results are delicately sensitive to the parameters of the prescribed ripples, which owing to uncertainty about the evolving, two-dimensional state of the ripples surrounding the transition to the turbulence, prevents unambiguous conclusions about CL validity.
In section~\ref{discussion}, we argue that this sensitivity implies that CL is ``incomplete'' because it does not also predict the \textit{response} of the wave field to the currents and turbulence beneath.

\section{Laboratory experiments of winds rising over calm water}
\label{lab}

This paper uses an experiment similar to those reported by \cite{melville1998laboratory} and \cite{veron2001experiments}
The experiment was conducted in the 42-m long, 1-m wide, 1.25-m high wind-wave-current tank at the Air-Sea Interaction Laboratory of the University of Delaware, and used a computer-controlled recirculating wind tunnel to accelerate a turbulent airflow to 10 $\mathrm{m \, s^{-1}}$ over 65 s.
The water depth was maintained at 0.71 m and observations were collected at a fetch of 12 m.
An artificial wave-absorbing beach dissipated wave energy and eliminate wave reflections at the downwind end of the tank.

\subsection{Laser-induced fluorescence (LIF) observations}

The evolution of initially-surface-concentrated dye was observed with a Laser-Induced Fluorescence (LIF) system.
Images were acquired with a CCD camera (Jai TM4200CL, 2048 × 2048 pixels) equipped with an 85 mm Canon EF lens focused at the air-water interface.
Illumination was provided by a thin 3 mm thick laser light sheet generated by a pulsed dual-head Nd-Yag laser (New Wave Research, 120 mJ/pulse, 3–5 ns pulse duration).
The laser light illuminated a thin layer of fluorescent dye carefully applied to the water surface prior to each experiment.
Observations were conducted with the light sheet in both along-wind and transverse directions. % (see figure \ref{fig:experiment}).
The LIF camera collected images at a 7.2Hz frame rate and with a field of view of 11.6 x 11.6 cm in the along-wind configuration, and 13.9 x 13.9 cm in the transverse direction.
An edge detection algorithm based on local variations of image intensity gradients, computed by kernel convolution was used to identify the location of the surface in the LIF images \citep{Buckley:2017}.

%\begin{figure}
%    \centering
%    \includegraphics[width = 0.8\textwidth]{experiment_v2.png}
%    \caption{Diagram of the laboratory setup, showing (a) the depth, along-wind section imaged by Laser Induced Fluoresence (LIF) and (b) the depth, cross-wind LIF section.}
%    \label{fig:experiment}
%\end{figure}

\subsection{Thermal Marking Velocimetry (TMV)}

In addition to LIF, we employed Thermal Marking Velocimetry (TMV), as developed by \cite{veron2001experiments} and \cite{veron2008wave}, to measure the surface velocity by tracking laser-generated Lagrangian heat markers in thermal imagery of water surfaces.
In the present experiment, infrared images of the surface were captured by a 14-bit, $640 \times 512$ quantum well infrared photodetector (QWIP --- 8.0-9.2$\mu$m) FLIR SC6000 infrared camera operated at a 43.2 Hz frame rate, with an integration time of 10 ms, and a stated rms noise level below 35 mK.
After image correction to account for the slightly off-vertical viewing angle of the imager, the resulting image sizes were $24.6 \times 24.6$ cm.

The infrared imager is sensitive enough to detect minute, turbulent temperature variations in the surface thermal skin layer \citep{jessup1997defining, Zappa2001, veron2001experiments, sutherland2013field}.
It thus easily detects active weakly-heated markers, generated by a 60 W air-cooled $\mathrm{CO_2}$ laser (Synrad Firestar T60) equipped with an industrial marking head (Synrad FH Index) and two servo-controlled scanning mirrors programmed to lay down a pattern of 16 spots with 0.8 cm diameter and at a frequency of 1.8 Hz.

The spatially averaged surface velocity was estimated by tracking the geometric centroid of these Lagrangian heat markers for approximately one second.
Both Gaussian interpolation (which has sub-pixel resolution due to the Gaussian pattern of the laser beam) and a standard cross-correlation technique yielded similar estimates for the surface velocity.

\subsection{Summary of experimental results}

\begin{figure}
    \centering
    \includegraphics[width = 0.75\textwidth]{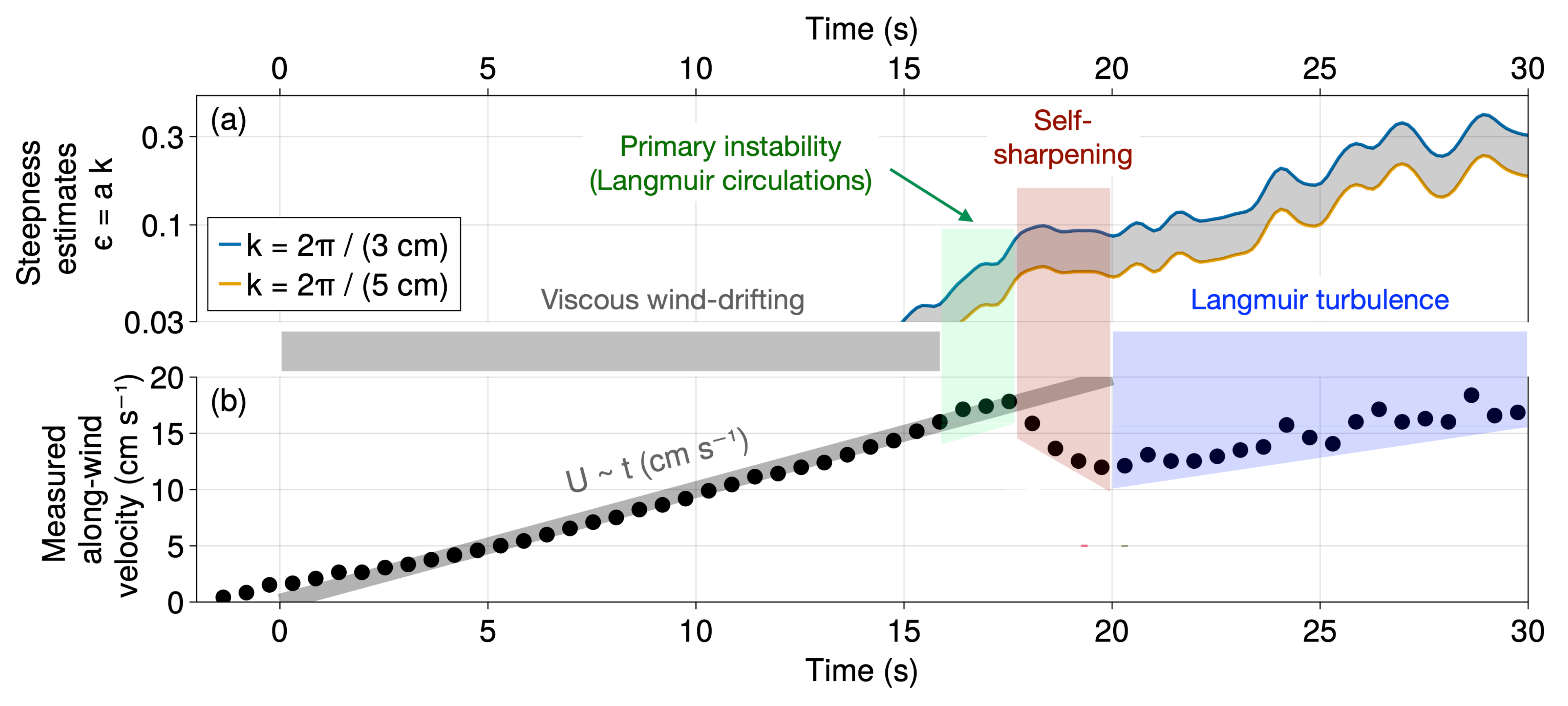}
    \caption{Summary of laboratory measurements: (a) an estimate of the steepness of the surface wave field that combines amplitude data with wavenumbers $k = 2 \pi / 3 \, \mathrm{cm^{-1}}$ and $k = 2 \pi / 5 \, \mathrm{cm^{-1}}$; (b) average surface along-wind velocity.}
    \label{lab-data-summary}
\end{figure}

Figure~\ref{lab-data-summary} summarizes the experimental results.
Figure~\ref{lab-data-summary}a shows a time series of estimated wave steepness, and figure~\ref{lab-data-summary}b shows the measured average surface velocity using TMV.
The thick gray line in figure~\ref{lab-data-summary} plots $A t$, where $A = 1 \, \mathrm{cm \, s^{-1}}$ showing that the surface current increases linearly in time.
(The time axis for laboratory measurements is adjusted to meet this line, which constitutes a definition of ``$t = 0$''.)
Following \cite{veron2001experiments}, figures~\ref{lab-data-summary}a--b divide the development of the waves and currents into four stages:

\begin{enumerate}
\item \textit{Viscous acceleration, $t=0$--$16$ s.} In the first stage, viscous stress between the accelerating wind and water accelerates a shallow, laminar, viscous wind-drift layer.
\item \textit{Wave-catalyzed ``Langmuir'' shear instability, $t=16$--$18$ s.} At $t \approx 16$ s, detectable capillary ripples appear.
A wave-catalyzed shear instability --- which obey the same dynamics as ``Langmuir circulation'', which often refers to much larger scale motions in the ocean surface boundary layer \cite{craik1976rational} --- immediately starts to develop and grow in the wind-drift layer.
\item \textit{Self-sharpening, $t=18$--$20$ s.} When the shear instability reaches finite amplitude, nonlinear amplification due to perturbation self-advection sharpens the instability features into narrow jets and downwelling plumes.
\item \textit{Langmuir turbulence, $t > 20$ s.} The self-sharpened circulations develop significant three-dimensional characteristics and transition to fully developed Langmuir turbulence.
\end{enumerate}

\section{A wave-averaged model for the transition to turbulence in wind-drift layers}
\label{theory}

The main purpose of this paper is to build a model for the four-stage evolution of the wind-drift layer, focusing on the dynamics after ripple inception.

\subsection{Viscous acceleration}

As the wind starts to accelerate, viscous stress across the air-water interface drives a laminar wind-drift current.
The thick gray line in figure~\ref{lab-data-summary} indicates that the average surface velocity nearly obeys
\beq
U(z=0, t) \defn U_0(t) = A \, t \com
\eeq
where $A \approx 1 \, \mathrm{cm \, s^{-2}}$.
\cite{veron2001experiments} point out that the viscous stress consistent with linear surface current acceleration is
\beq \label{time-dependent-stress}
\boldsymbol{\tau}(t) = \alpha \sqrt{t} \, \bxh
\eeq
where $\boldsymbol{\tau}$ is the downwards kinematic stress across the air-water interface, $\bxh$ is the along-wind direction ($\byh$ and $\bzh$ are the cross-wind and vertical directions), and $\alpha \approx 1.2 \times 10^{-5} \, \mathrm{m^2 \, s^{-5/2}}$ produces $A = \alpha \sqrt{4 / \pi \nu} \approx 1 \, \mathrm{cm \, s^{-1}}$ given the kinematic viscosity of water, $\nu = 1.05 \times 10^{-6} \, \mathrm{m^2 \, s^{-1}}$.
The laminar viscous solution beneath \eqref{time-dependent-stress} is \citep{veron2001experiments}
\beq \label{mean-flow}
U(z, t) = U_0(t) \left [ \left ( 1 + \delta^2 \right ) \mathrm{erfc} \left ( - \tfrac{\sqrt{2}}{2} \delta \right ) + \delta \sqrt{\tfrac{2}{\pi}} \exp \left ( - \tfrac{1}{2} \delta^2 \right ) \right ]
\quad \text{where} \quad
\delta \defn \frac{z}{\sqrt{2 \nu t}} \per
\eeq
Viscous acceleration continues until gravity-capillary ripples appear at the air-water interface when $t \approx \tilde t \defn 16$ seconds and thus $U_0(t) \approx \tilde U_0 \defn 16 \, \mathrm{cm \, s^{-1}}$.
%We note that this rippling instability is not described by linear stability theory \citep{young2014generation}.

\subsection{Instability of the wind-drift layer catalyzed by incipient capillary ripples}
\label{stage-two}

As soon as ripples appear on the water surface, a second, slower, non-propagating instability begins to grow.
Remarkably, this second instability is catalyzed by and therefore requires the presence of capillary ripples: for example \cite{veron2001experiments} show that instability and turbulence are suppressed if ripple generation is suppressed by layering a surfactant on the water surface.

Interestingly, ripple inception is not explained by linear theory \citep{young2014generation}.
We neverthless assume that the ensuing dynamics are described by the weakly-nonlinear and wave-averaged ``Craik-Leibovich'' momentum equation.
This ansatz is supported by the data in figure~\ref{lab-data-summary}a, which plots an estimate of the wave steepness $\epsilon \defn a k$, where $a$ is the surface wave amplitude (readily measured in the lab) and $k \approx 2\pi/3 \, \mathrm{cm^{-1}}$--$2 \pi/5 \, \mathrm{cm^{-1}}$ is the gravity-capillary wavenumber for ripples with wavelengths between 3 and 5 cm (slightly larger than the minimum gravity-capillary wavelength 1.7 cm).
Apparently, by the time ripples reach detectable amplitudes they have small slopes with $\epsilon < 0.1$.

The wave-averaged Craik-Leibovich equation \citep{craik1976rational} formulated in terms of the Lagrangian-mean momentum $\buL$ of the wind-drift layer is
\beq \label{momentum}
\buL_t + \left ( \buL \bcdot \bnabla \right ) \buL
    - \left ( \bnabla \times \buS \right ) \times \buL + \bnabla \pE = \nu \lap \buL - \nu \lap \buS + \buS_t \com 
\eeq
where $\buS$ is the Stokes drift of the field of capillary ripples and $\pE$ is the Eulerian-mean pressure.
The asymptotic derivation of the CL equation \eqref{momentum} requires $\epsilon \ll 1$.
% (Note that $\nu \lap \buS$ is a ``wave streaming'' term describing the redistribution of momentum during the viscous dissipation of the surface wave field \citep{longuet1953mass}.
% We include this term for completeness, but it has little effect on our results.\textcolor{red}{Nick: Doesn't this assume that the waves have super tiny amplitude compared to the viscous boundary layer depth, a scenario which is almost never true at the surface? See p. 539 of that paper, second paragraph})
%We limit our attention to solutions of \eqref{momentum} beneath steady wave fields with $\d_t \buS = 0$.
%describes a transfer of momentum from air to water during the forced growth or decay of surface waves \citep{wagner2021near}.
We require $\buL$ to be divergence-free \citep{vanneste2022stokes},
\beq \label{continuity}
\bnabla \bcdot \buL = 0 \per
\eeq
%(see \cite{vanneste2022stokes} and Wagner \textit{et al.} 2021) \nocite{wagner2021near}. in the case that $\d_t \uS(z, t) \ne = 0$ changes in time.
The Stokes drift associated with monochromatic capillary ripples propagating in the along-wind direction $\bxh$ is
\beq \label{stokes-drift}
\buS(z, t) = \ee^{2 k z} \epsilon^2 c(k) \, \bxh \com
\qquad \text{where} \qquad 
c(k) = \sqrt{\frac{g}{k} + \gamma k}
\eeq
is the phase speed of gravity-capillary waves in deep water with gravitational acceleration $g = 9.81 \, \mathrm{m \, s^{-2}}$ and surface tension $\gamma = 7.2 \times 10^{-5} \, \mathrm{m^3 \, s^{-2}}$.
In all cases considered here, the Stokes drift \eqref{stokes-drift} is minuscule compared to the mean current $\buL \sim U$.
%This is probably a robust feature of the transition to turbulent in wind-drift layers, since both $c \approx U$ and $\epsilon \ll 1$ shortly after ripple inception, implying that $\buS \ll U$ in \eqref{stokes-drift}.

% Our first question is whether the wind-drift layer is susceptible to further instability following the initial instability that generates capillary ripples.
% \cite{holm1996ideal} shows that the necessary condition for shear instability in the context of the wave-averaged equations \eqref{momentum}--\eqref{continuity} with $\nu=0$, steady waves, and steady currents, is
% \beq \label{necessary-condition}
% \frac{\mathrm{d}^2}{\mathrm{d} z^2} \left ( \buL - \buS \right ) = 0 \com
% \eeq
% at some level $z$.
% We find that \eqref{necessary-condition} is always satisfied for the shear flow $\buL = U(z, t) \, \bxh$ in \eqref{mean-flow} with the Stokes drift \eqref{stokes-drift} for $\epsilon > 0$ and $t >0$.

% We need more detail than just the necessary condition for instability.
We invoke the power method \citep{constantinou2015formation} to investigate the stability of the wind-drift layer immediately following ripple generation.
%For this we invoke the ``power method'' \citep{constantinou2015formation} to diagnose the fastest growing unstable mode (if any exists) of the shear current at wave inception beneath gravity-capillary ripples with wavenumber $k = 2\pi / 3 \, \mathrm{cm^{-1}}$, given a range of wave amplitudes.
In our implementation of the power method we extract the fastest growing mode perturbing the steady shear flow $\tilde U(z)$, such that
\beq \label{linear-stability-current}
\buL(y, z, t) = \tilde U(z) \, \bxh + \bu(y, z, t) \com 
\eeq
where $\tilde U(z) \defn U(z, \tilde t)$ represents the wind-drift profile ``frozen'' at $\tilde t = 16$ s and $\bu = (u, v, w)$ is the perturbation velocity.
Inserting \eqref{linear-stability-current} into \eqref{momentum}--\eqref{continuity}, introducing a streamfunction $\psi$ with the convention $(v, w) = (-\psi_z, \psi_y)$, and neglecting terms that depend only on the mean flow $U$ or $u^S$ yields the two-dimensional system
\begin{gather}
u_t + \J(\psi, u) + \Omega \, \psi_y = \nu \lap u \com \label{linearized-x-momentum} \\
\lap \psi_t + \J(\psi, \lap \psi) + \uS_z u_y = \nu \lap^2 \psi \label{linearized-x-vorticity} \com
\end{gather}
where $\lap \psi = w_y - v_z$ is the $x$-component of the perturbation vorticity, $\J(a, b) = a_y b_z - a_z b_y$ is the Jacobian operator, and $\Omega = \tilde U_z - \uS_z$ is the Eulerian-mean shear --- or, as we prefer, the mean, \textit{total} cross-wind vorticity $\bnabla \times \left ( \tilde U \bxh - \buS \right ) = \Omega \byh$.
The power method isolates the fastest growing linear modes of \eqref{linearized-x-momentum}--\eqref{linearized-x-vorticity} by iterating over short integrations of the fully nonlinear CL equations \eqref{momentum}--\eqref{continuity} from $t = \tilde t$ to $t = \tilde t + \Delta t$ and repeatedly downscaling $\bu$ via \eqref{linear-stability-current} to ensure each integration has essentially linear dynamics.

\begin{figure}
    \includegraphics[width = 1\textwidth]{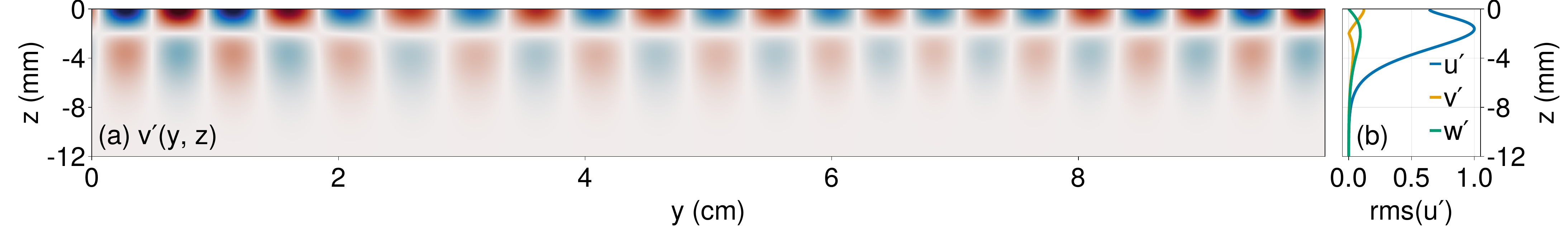}
    \caption{
    Structure of the most unstable eigenmode for wave steepness $\epsilon = 0.1$ in a $10 \times 5$ cm domain in $(y, z)$.
    (a) Structure of the cross-wind perturbation $v'(y, z)$ for wave steepness $\epsilon = 0.1$ in a $10 \times 5$ cm domain in $(y, z)$, (b) root-mean-square ($y$-averaged) perturbation profiles}.
    \label{fig:langmuir-mode}
\end{figure}

\begin{figure}
    \centering
    \includegraphics[width = 0.8\textwidth]{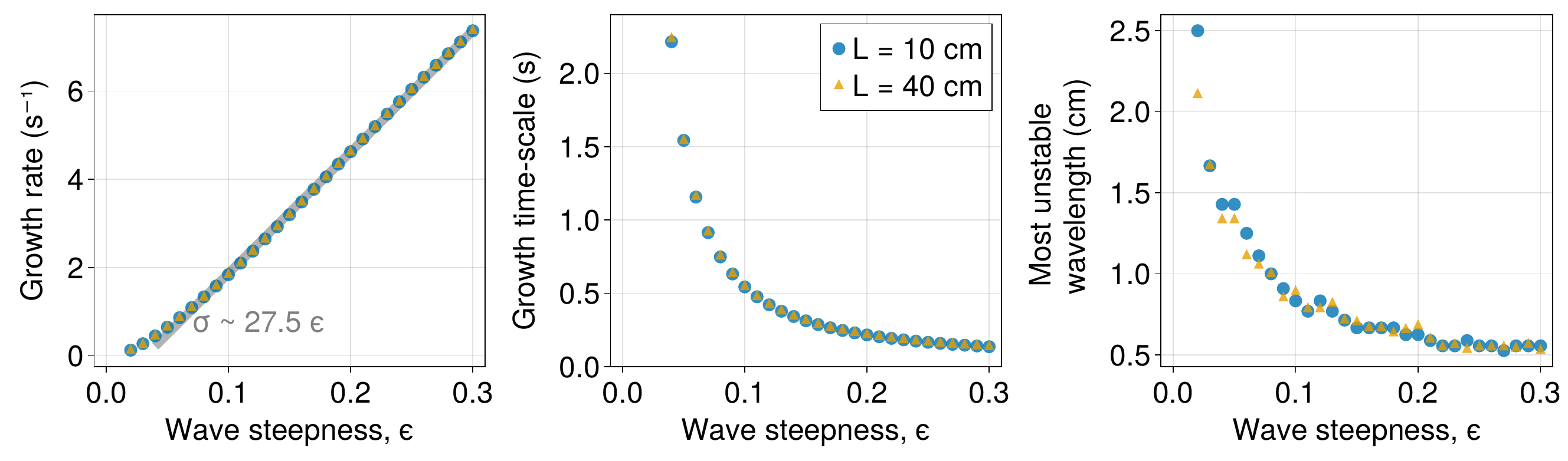}
    \caption{(a) Growth rate, (b) growth time-scale (inverse of the growth rate), and (c) wavelength of the most unstable mode of a wave-catalyzed instability of the wind-drift layer ``frozen'' at $\tilde t = 16$ seconds calculated using the power method \citep{constantinou2015formation} in two domains 10 cm and 40 cm wide.}
    \label{fig:linear-stability}
\end{figure}

Figure~\ref{fig:langmuir-mode} shows the structure of the most unstable mode for $\epsilon = 0.1$.
The mode is necessarily sinusoidal in the horizontal because \eqref{linearized-x-momentum}-\eqref{linearized-x-vorticity} is horizontally-uniform and has alternating upwelling and downwelling regions, with a much larger streamwise component $u'$ than cross-stream and vertical components $v'$ and $w'$.
Figure~\ref{fig:linear-stability} shows the results of a parameter sweep from $\epsilon = 0.04$ to $\epsilon = 0.3$, illustrating how increasing $\epsilon$ leads to faster growth rates and decreasing wavelengths of the fastest growing mode.
We recover \cite{veron2001experiments}'s experimental result that shear instability does not occur without surface waves.
% \textcolor{red}{The wavelengths that come out of our calculations are a lot smaller than the surface streak spacings observed by \cite{melville1998laboratory} --- which vary between 3 and 8 cm --- why is that? Nick: The fastest growing mode of steep surface waves is oblique (not normal) to the direction of wave propagation so if it is the 2d horseshoe pattern setting that transverse scale, it is plausible it's longer than the wavelength of the waves.}

Note that the kinetic energy source for growing perturbations is the mean shear $\tilde U(z)$ and there is no energy exchange between perturbations and the surface wave field within the context of the wave-averaged equations \eqref{momentum}--\eqref{continuity}.
To see this, consider that \eqref{momentum}--\eqref{continuity} conserves total kinetic energy $\int \tfrac{1}{2} | \buL |^2 \di V$ when $\nu=0$ and $\d_t \buS = 0$ (viscous stress is negligible).
%; the Reynolds number at wave inception based on the surface velocity and wind-drifted depth is $Re = A \sqrt{\tilde t^3 / \nu} \approx 625$).
This is why we characterize the shear instability as ``wave-catalyzed'': while the presence of waves is necessary for instability, and while the instability growth rate is strongly affected by wave amplitude, the kinetic energy of the growing perturbation is derived solely from the mean shear.

\subsection{Self-sharpening circulations with jets and plumes}

When the wave-catalyzed shear instability reaches finite amplitude, it begins to self-sharpen, producing narrow along-wind jets and downwelling plumes.
The sharpening --- but still two-dimensional --- plumes then transport a measureable amount of mean momentum downwards before becoming unstable to three-dimensional perturbations and thereby transitioning to fully-developed turbulence.
This nonlinear sharpening and depletion of the average near-surface momentum occurs between 17.5--20 seconds, as evidenced by the red shaded region in figure~\ref{lab-data-summary}b.

To simulate the second wave-catalyzed instability through finite amplitude and toward transition to turbulence, we propose a simplified model based on the wave-averaged equations \eqref{momentum}--\eqref{continuity} with two main components representing \textit{(i)} the capillary ripples and \textit{(ii)} the initial condition at $\tilde t = 16$ seconds.
We model the evolving capillary ripples as steady, monochromatic surface waves with wavenumber $k = 2\pi / 3 \, \mathrm{cm^{-1}}$.
We model the condition of the wave tank at $t = 16$ seconds with
\beq
\buL \, |_{t=\tilde t} = \tilde U(z) \bxh + \mathbb{U}' \, \boldsymbol{\Xi}(x, y, z) \com
\eeq
where $\mathbb{U}'$ is an initial noise amplitude and $\boldsymbol{\Xi}$ is a vector whose components are normally-distributed random numbers with zero mean and unit variance.

Through experimentation, we find that the instability and transition to turbulence are only weakly sensitive to the wavenumber $k$.
The tuning parameters of our model are therefore \textit{(i)} the amplitude of the initial perturbation $\mathbb{U}'$, and \textit{(ii)} the wave steepness $\epsilon$.
We discuss possible interpretations for $\mathbb{U}'$ in section~\ref{discussion}.

We also simulate the evolution of dye concentration $d$ via
\beq \label{tracer}
\d_t \theta + \buL \bcdot \bnabla \theta = \kappa \triangle \theta \com 
\eeq
with molecular diffusivity $\kappa = 10^{-7} \, \mathrm{m^2 \, s^{-1}}$, the smallest we can reasonably afford computationally.
(The correspondance between $\theta$ and rhodamine is imperfect because the molecular diffusivity of rhodamine is $\kappa = 10^{-9} \, \mathrm{m^2 \, s^{-1}}$.)
We initialize $\theta$ with a $\delta$-function at the surface.
%, since the thickness of the diffusing layer of rhodamine $\sqrt{\kappa \tilde t} \approx 0.1$ mm at $\tilde t = 16$ seconds, close to the grid scale.
We integrate \eqref{momentum}--\eqref{continuity} and \eqref{tracer} given \eqref{stokes-drift} using Oceananigans (see \citealt{ramadhan2020oceananigans} and \citealt{wagner2021near}) with a second-order staggered finite volume method on a single A100 GPU in a $10 \times 10 \times 5$ cm domain in $(x, y, z)$ with 0.13 mm regular spacing in $x, y$ and variable spacing in $z$ with $\min(\Delta z) \approx 0.26 \, \mathrm{mm}$, corresponding to $768 \times 768 \times 512$ finite volume cells.

\begin{figure}
    \includegraphics[width = 1\textwidth]{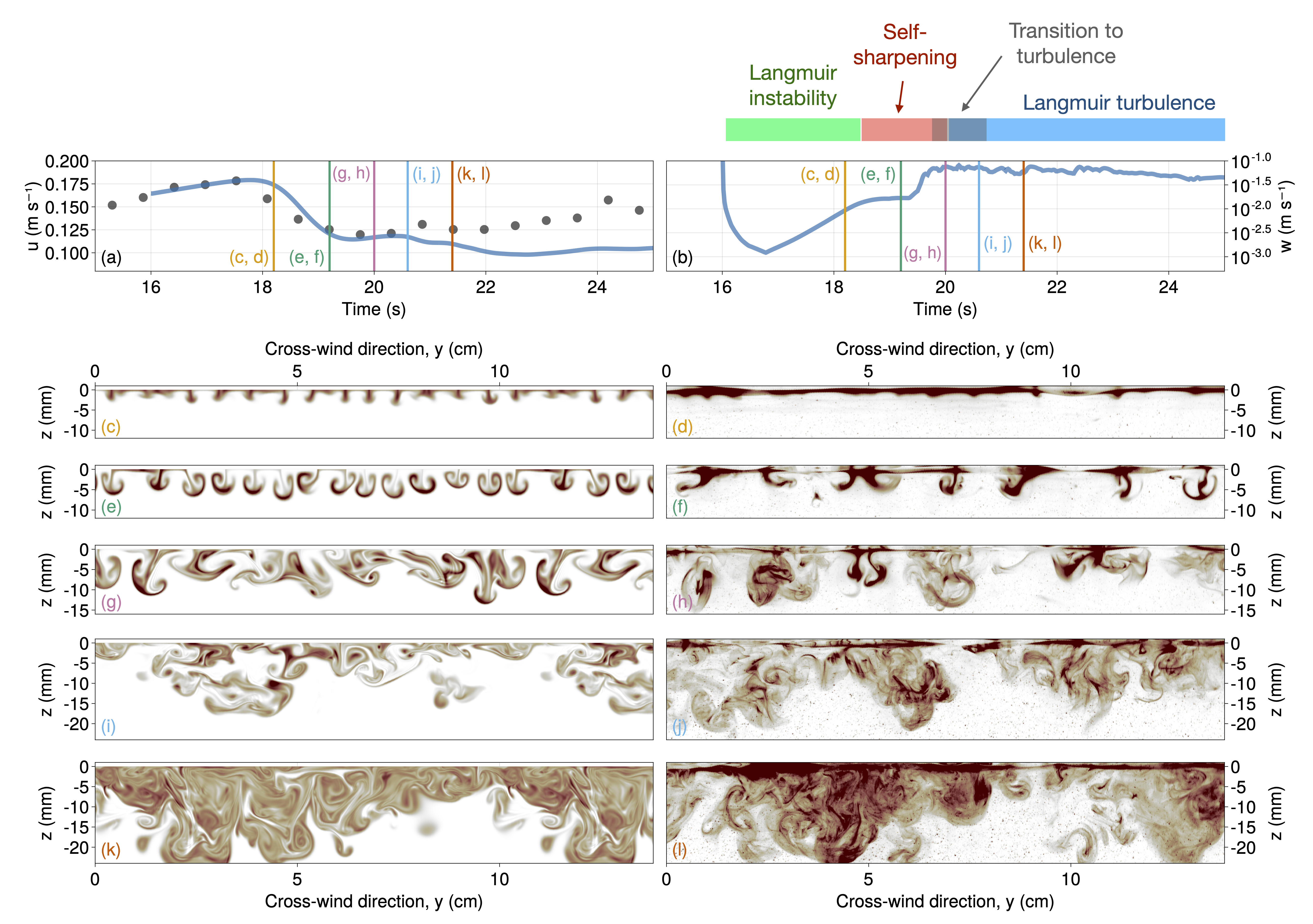}
    \caption{(a) Maximum horizontal velocity $u$. The grey dots are from the laboratory experiments while the blue line shows the numerical experiments. (b) Vertical velocity.  (c-l) Evolution of the wind-drift layer after the emergence of capillary ripples.}
    \label{fig:self-sharpening}
\end{figure}

The results are visualized in figure~\ref{fig:self-sharpening}.
Figure~\ref{fig:self-sharpening}a compares the average surface velocity diagnosed from the simulation with the laboratory measurements presented in figure~\ref{lab-data-summary}b.
Figure~\ref{fig:self-sharpening}b plots the maximum absolute vertical velocity, $\max|w^L|$.
Figures~\ref{fig:self-sharpening}c--l compare the simulated dye concentration on $(y, z)$ slices with LIF measurements of rhodamine from the laboratory experiment, showing the particularly good qualitative agreement between simulated and measured dye features.
Visualizations are shown at $t=18.1$, 19.3, 20.0, 20.7, and 21.7 seconds.% for figures~\ref{fig:self-sharpening}c-d, e-f, g-h, i-j, and k-l respectively.
At $t = 18.1$ s (figures~\ref{fig:self-sharpening}c and d), the sharpened plumes have only just started advect appreciable amounts of dye.
At $t = 19.3$ seconds (figures~\ref{fig:self-sharpening} e and f), the plumes are beginning to roll up into two-dimensional mushroom-like structures.
(Note the small, unexplained discrepancy between simulated and measured average surface velocities around $18.1 < t < 19.3$.)
At $t = 20.0$ seconds and $t = 20.7$ (figures~\ref{fig:self-sharpening}g, h, i, and j) three-dimensionalization and transition to turbulence are underway.
At $t = 21.7$ both the simulated and measured dye concentrations appear to be mixed by three-dimensional turbulence.

\begin{figure}
    \centering
    \includegraphics[width = 0.8\textwidth]{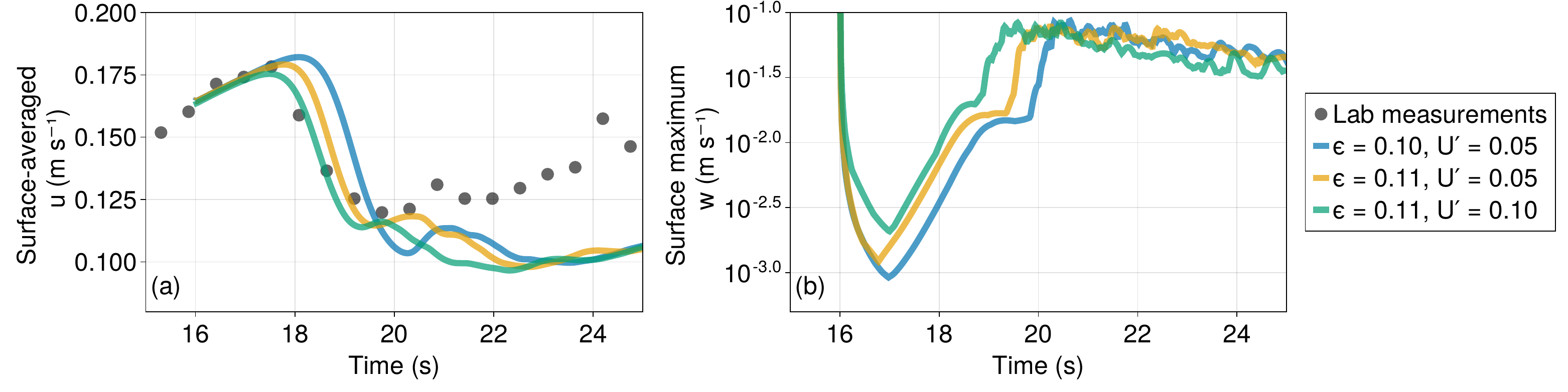}
    \caption{Sensitivity of (a) surface-averaged along-wind velocity $u$ and (b) maximum vertical velocity to the specification of the surface wave field and initial perturbations. %Figure~\ref{fig:self-sharpening} uses $\epsilon = 0.11$ and $\mathbb{U}'=5 \, \mathrm{cm \, s^{-1}}$. With $\epsilon=0.11$, $\mathbb{U}'=10 \, \mathrm{cm \, s^{-1}}$ the boundary layer evolution proceeds more quickly than the experiment, as evidenced by dye visualization (not shown).
    %Roughly 20\% changes in wave amplitude to either $\epsilon=0.09$ or $\epsilon=0.13$ produce too-slow or too-fast boundary layer evolution.
    %We estimate thus that matching the evolution of the dye simultaneously fixes the wave field magnitude to within 10\% and the initial perturbation to within 50\%. 
    }
    \label{fig:sensitivity}
\end{figure}

Figure~\ref{fig:sensitivity} illustrates the sensitivity of the wave-averaged model to amplitude of the specified surface ripples and to the amplitude of the random initial perturbation.
Figures~\ref{fig:sensitivity}a and b plot the surface-averaged along-wind velocity $u$ and maximum vertical velocity $w$ for two wave amplitudes $\epsilon = 0.1$ and $\epsilon = 0.11$, and for two initial perturbaiton ampltidues $\mathbb{U}'=5 \, \mathrm{cm \, s^{-1}}$ and $\mathbb{U}'=10 \, \mathrm{cm \, s^{-1}}$.
The dependence of the maximum vertical velocity is the most evocative: doubling the initial perturbation shortens the self-sharpening phase (in which the maximum vertical velocity in figure~\ref{fig:sensitivity}b flattens before increasing sharply during the transition to turbulence) by a factor of five.
Of the three cases plotted in figure~\ref{fig:sensitivity}, only $\epsilon = 0.11$ and $\mathbb{U}' = 5 \, \mathrm{cm \, s^{-1}}$ yield the satisfying agreement depicted in figure~\ref{fig:self-sharpening}.

\subsection{Langmuir turbulence}

\begin{figure}
    \centering
    \includegraphics[width = 1.0\textwidth]{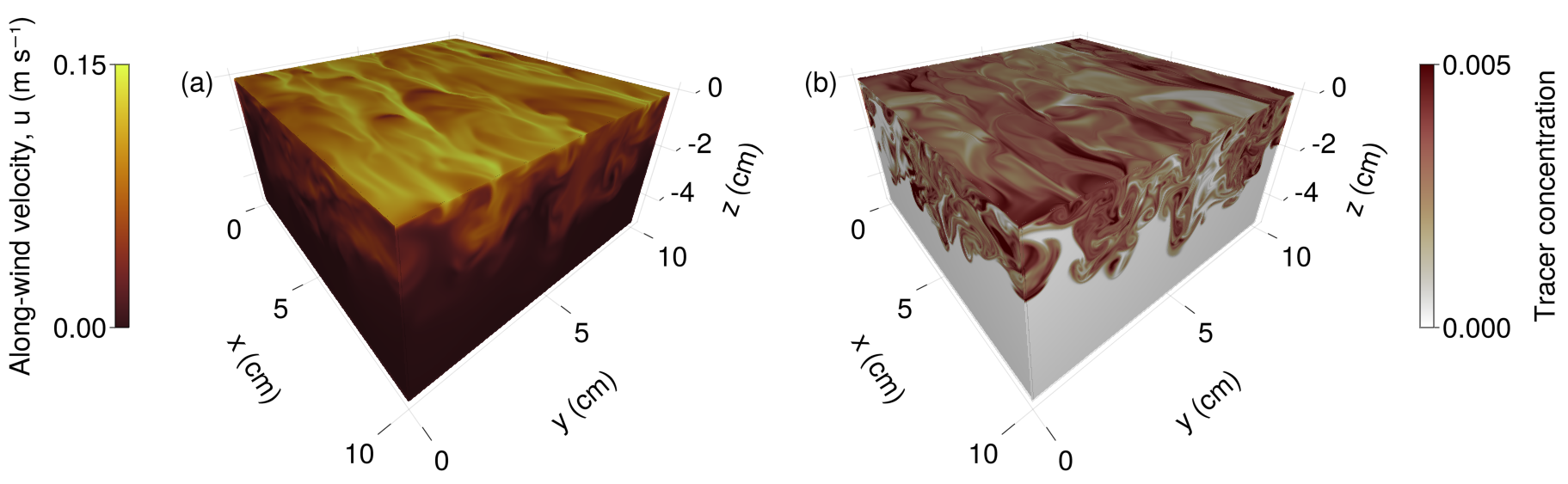}
    \caption{(a) Simulated $x$-momentum and (b) dye concentration at $t = 23.4$ seconds, showing the streaks and jets that characterize Langmuir turbulence.}
    \label{fig:langmuir}
\end{figure}

Following three-dimensionalization, momentum and dye are rapidly mixed to depth.
Figure~\ref{fig:langmuir} visualizes (a) the $x$-momentum and (b) dye concentration at $t=23.4$ seconds, showing how the flow is organized into narrow along-wind streaks and broader downwelling regions --- classic characteristics of Langmuir turbulence \citep{sullivan2010dynamics}.

\begin{figure}
    \centering
    \includegraphics[width = 0.8\textwidth]{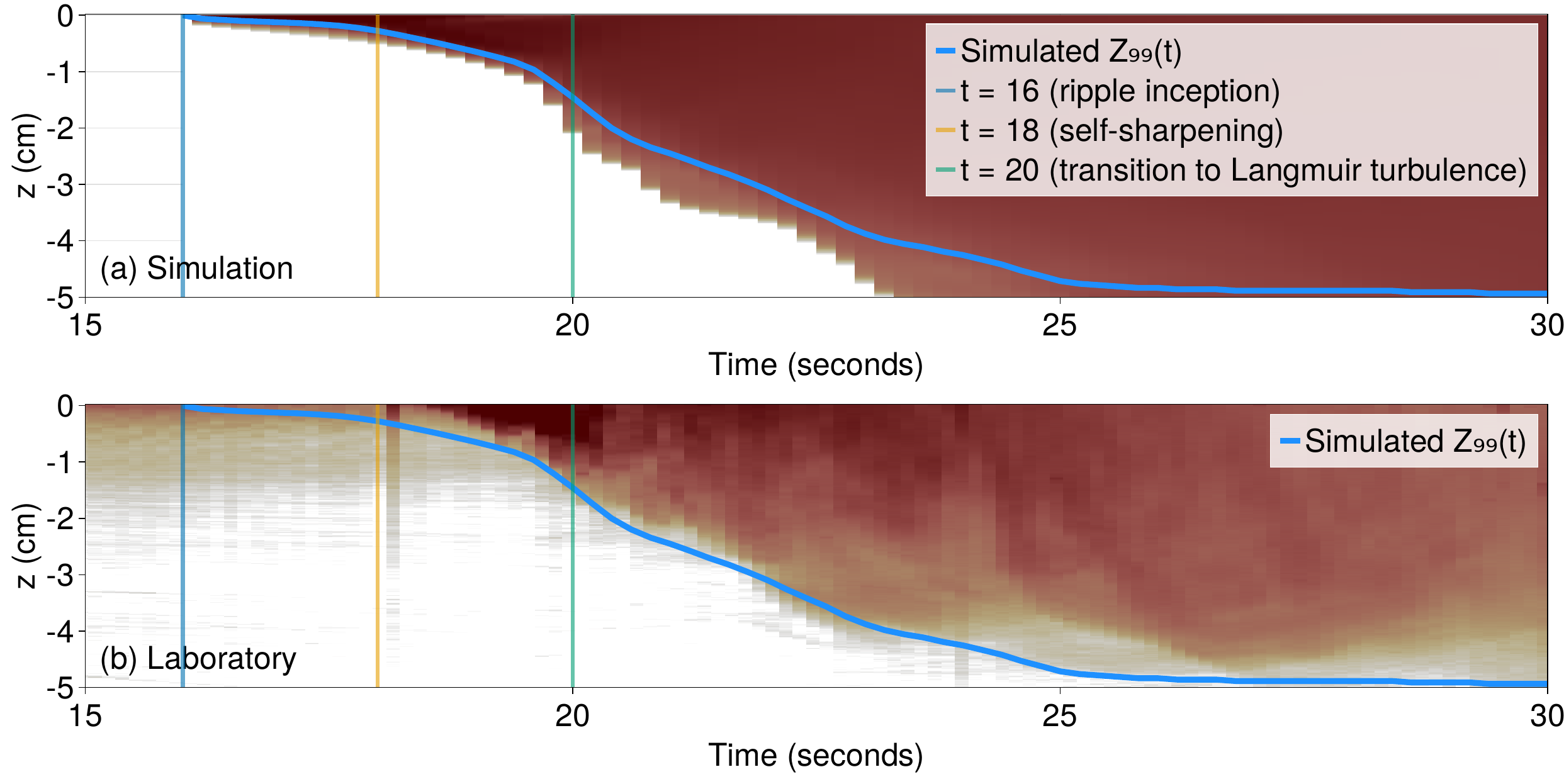}
    \caption{Visualization of mixing rates during measured and simulated wave-catalyzed instability via depth-time $(z, t)$ diagrams of horizontally-averaged in (a) simulations and (b) LIF-measurements.
    The blue lines denotes the depth above which 99\% of the simulated dye resides; (b) suggests that the simulations overpredict dye mixing rates.}
    %(c) Comparison of centroids of the simulated dye concentration in (a) and the measured dye concentration in (b). Though (c) suggests similarity between the simulated and measured rates of dye mixing, inspecting (a) and (b) shows that the simulations overpredict the rate of dye mixing in the laboratory.}
    \label{fig:hovmoller}
\end{figure}

In figure~\ref{fig:hovmoller}, we compare the rate which dye is mixed to depth in the simulations versus measured by LIF in the lab.
Figure~\ref{fig:hovmoller}a shows the simulated horizontally-averaged tracer concentration in the depth-time $(z, t)$ plane, while figure~\ref{fig:hovmoller}b shows a corresponding laboratory measurements extracted from LIF measurements in the $(y, z)$-plane.
In figures~\ref{fig:hovmoller}a and b, a light blue line shows the height $z_{99}(t)$ defined as the level above which 99\% of the \textit{simulated} tracer concentration resides,
\beq
\int_{z_{99}(t)}^0 \theta \di z = 0.99 \int_{-H}^0 \theta \di z \per
\eeq
Using $z_{99}(t)$ to compare the tracer mixing rates exhibited in figures~\ref{fig:hovmoller}a and b, we conclude that the simulations provided a qualitatively accurate prediction of dye mixing rates.
If anything, the simulation \textit{overpredicts} the dye mixing rate --- but the data probably does not warrant more than broad qualitative conclusions.

\section{Discussion}
\label{discussion}

This paper describes a wave-averaged model for the evolution of wind-drift layers following the inception of capillary ripples.
The wave-averaged model predicts that, following ripple incpetion, the wind-drift layer is immediately susceptible to the growth of a second, slower, ``wave-catalyzed'' instability.
Wave-averaged simulations show that the evolution of the wave-catalyzed instability from initial growth through transition to turbulence is sensitive not only to the amplitude of the surface ripples, but also to the amplitude of the perturbations that seed the growth of the second-instability and also destabilize the jets and plumes during the transition to turbulence.

We model the seeding and destabilizing perturbations modeled as random velocity fluctuations imposed at the time of ripple inception.
However, in wind-drift layers in the laboratory or natural world, perturbations may be continuously introduced both by turbulent pressure fluctuations in the air and, perhaps, by inhomogeneities in the ripple field (see figures 3 and 15 in \citealt{veron2001experiments}).
The natural prevalance of the striking coherent structures observed in the laboratory and visualized in figure~\ref{fig:self-sharpening} is therefore uncertain.

One of the original goals of this work was to probe the potential weaknesses of the wave-averaged Craik-Leibovich (CL) equation, which are widely used for process studies and paramterization of ocean surface boundary layer turbulence \citep{sullivan2010dynamics}.
In particular, the derivation of the CL equation requires the restrictive assumptions that
\textit{(i)} the surface wave field is nearly linear; and
\textit{(ii)} the timescale of the turbulence is much longer than the time scale of the waves,
But despite qualitative success, firm conclusions about CL validity prove elusive due to the strong sensitivity of our results to ripple amplitude --- which is evolving and two-dimensional in the laboratory experiments rather than uniform and steady, as in our model.
This sensitivity, together with \cite{veron2001experiments}'s observations that the ripple field is refracted and organized by turbulence in the wind-drift layer, suggests that two-way wave-turbulence coupling is important and should be described in a ``predictive'' theory for turbulent boundary layers affected by surface waves.
Further progress requires not just theoretical advances to couple wave evolution with the CL equations, but also experiments that obtain more precise two-dimensional measurements of the evolution of the capillary ripples.

\begin{acknowledgments}
\textit{Acknowledgements.}
We acknowledge stimulating discussions with Bill Young and Navid Constantinou.
\end{acknowledgments}

\bibliographystyle{jfm}
\bibliography{refs}

\end{document}